\documentclass[9pt, conference]{IEEEtran}
\IEEEoverridecommandlockouts

\usepackage{cite}
\usepackage{amsmath,amssymb,amsfonts}
\usepackage{algorithmic}
\usepackage{graphicx}
\usepackage{textcomp}
\usepackage{xcolor}

\newcommand{\degree}{\ensuremath{^\circ}\xspace}
\newcommand{\ignorethis}[1]{}

\makeatletter
\DeclareRobustCommand\onedot{\futurelet\@let@token\@onedot}
\def\@onedot{\ifx\@let@token.\else.\null\fi\xspace}

\makeatother

\definecolor{MyDarkBlue}{rgb}{0,0.08,1}
\definecolor{MyDarkGreen}{rgb}{0.02,0.6,0.02}
\definecolor{MyDarkRed}{rgb}{0.8,0.02,0.02}
\definecolor{MyDarkOrange}{rgb}{0.40,0.2,0.02}
\definecolor{MyPurple}{RGB}{111,0,255}
\definecolor{MyRed}{rgb}{1.0,0.0,0.0}
\definecolor{MyGold}{rgb}{0.75,0.6,0.12}
\definecolor{MyDarkgray}{rgb}{0.66, 0.66, 0.66}

\newcommand{\myparagraph}[1]{\vspace{2pt}\noindent\textbf{#1}.\;}

\newcommand{\topology}{topology\xspace}

\newcommand{\topologies}{topologies\xspace}

\newcommand{\NAME}{GCN-RL Circuit Designer\xspace}

\usepackage{color,xcolor}
\usepackage{epsfig}
\usepackage{graphicx}

\usepackage{adjustbox}
\usepackage{array}
\usepackage{booktabs}
\usepackage{colortbl}
\usepackage{float,wrapfig}
\usepackage{hhline}
\usepackage{multirow}
\usepackage[labelfont={bf},font={small}]{caption}
\usepackage{subcaption} % issues a warning with CVPR/ICCV format
\usepackage{amsfonts}
\usepackage{mathtools}
\usepackage{bm}
\usepackage{nicefrac}

\usepackage{changepage}
\usepackage{extramarks}
\usepackage{fancyhdr}
\usepackage{setspace}
\usepackage{soul}
\usepackage{xspace}
\usepackage{textcomp}

\usepackage{enumerate}
\usepackage{fancyhdr,graphicx,amsmath,amssymb}
\usepackage[ruled,vlined]{algorithm2e}

\def\BibTeX{{\rm B\kern-.05em{\sc i\kern-.025em b}\kern-.08em
    T\kern-.1667em\lower.7ex\hbox{E}\kern-.125emX}}

\makeatletter
 \let\old@ps@headings\ps@headings
 \let\old@ps@IEEEtitlepagestyle\ps@IEEEtitlepagestyle
 \def\confheader#1{%
 \def\ps@headings{%
 \old@ps@headings%
 \def\@oddhead{\strut\hfill#1\hfill\strut}%
 \def\@evenhead{\strut\hfill#1\hfill\strut}%
 }%
 \def\ps@IEEEtitlepagestyle{%
 \old@ps@IEEEtitlepagestyle%
 \def\@oddhead{\strut\hfill#1\hfill\strut}%
 \def\@evenhead{\strut\hfill#1\hfill\strut}%
 }%
 \ps@headings%
 }
 \makeatother

\confheader{%
 The 57th Design Automation Conference (DAC 2020)
 }

\begin{document}

\title{GCN-RL Circuit Designer: Transferable Transistor Sizing with Graph Neural Networks and Reinforcement Learning}

\author{
\\[-4.0ex]
\IEEEauthorblockN{Hanrui Wang$^1$, Kuan Wang$^1$, Jiacheng Yang$^1$, Linxiao Shen$^2$, Nan Sun$^2$, Hae-Seung Lee$^1$, Song Han$^1$}
\IEEEauthorblockA{\textit{$^1$Massachusetts Institute of Technology} \\ \textit{$^2$University of Texas at Austin}
\\[-4.0ex]
}

}

\maketitle

\begin{abstract}
Automatic transistor sizing is a challenging problem in circuit design due to the large design space, complex performance trade-offs, and fast technological advancements. Although there has been plenty of work on transistor sizing targeting on one circuit, limited research has been done on transferring the knowledge from one circuit to another to reduce the re-design overhead. In this paper, we present \NAME, leveraging reinforcement learning (RL) to transfer the knowledge  between different technology nodes and \topologies. Moreover, inspired by the simple fact that \emph{circuit is a graph}, we learn on the circuit \topology representation with graph convolutional neural networks (GCN). The GCN-RL agent extracts features of the \topology graph whose vertices are transistors, edges are wires. Our learning-based optimization consistently achieves the highest Figures of Merit (FoM) on four different circuits compared with conventional black-box optimization methods (Bayesian Optimization, Evolutionary Algorithms), random search, and human expert designs. Experiments on transfer learning between five technology nodes and two circuit \topologies demonstrate that RL with transfer learning can achieve much higher FoMs than methods without knowledge transfer. Our transferable optimization method makes transistor sizing and design porting more effective and efficient.
\end{abstract}


\section{Introduction}
Mixed-signal integrated circuits are ubiquitous. While digital designs can be assisted by the mature VLSI CAD tools~\cite{elfadel2018machine}, analog designs still rely on experienced human experts. It is demanding to have learning-based design automation tools.

Nonetheless, manual design is not an easy task even for seasoned designers due to the long and complicated design pipeline. Designers first need to analyze the topology and derive equations for the performance metrics. Since analog circuits have highly nonlinear properties, a large number of simplifications and approximations are necessary during the topology analysis. Based on all the equations, the initial sizes are calculated. Then, a large number of simulations for parameters fine-tuning are performed to meet the performance specifications. The whole process can be highly \emph{labor intensive} and \emph{time consuming} because of the large design space, slow simulation tools, and sophisticated trade-offs between different performance metrics.
Therefore, automatic transistor sizing is attracting more and more research interest in the recent years~\cite{lyu2018batch, liao2017parasitic, liu2010enhanced, learncircuits}.

\begin{figure}[t]
    \centering
    \includegraphics[width=\linewidth]{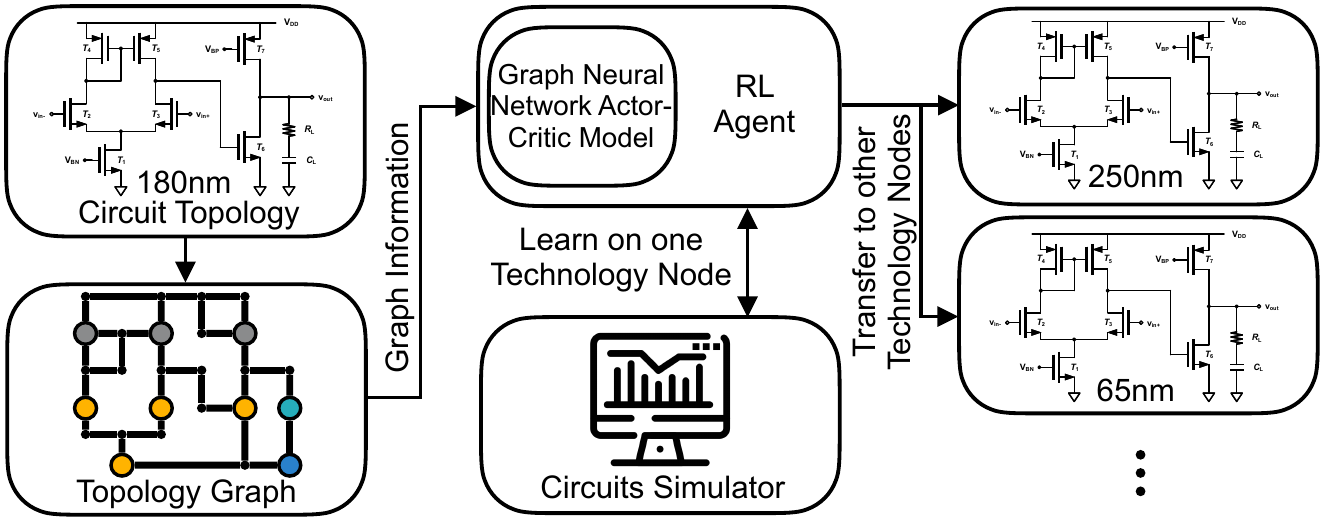}
    \caption{Graph Convolutional Neural Networks based Reinforcement Learning for Automatic Transistor Sizing.}
    \vspace{-10pt}
    \label{fig:teaser}
\end{figure}

With transistors rapidly scaling down, \emph{porting} existing designs from one technology node to another become a common practice. However, although much research efforts focus on transistor sizing for a single circuit, hardly any research has explored transferring the knowledge from one \topology to another, or from one technology node to another. In this work, we present \NAME (Figure \ref{fig:teaser}) to conduct the knowledge transfer. Inspired by the transfer learning ability of \emph{Reinforcement Learning} (RL), we first train a RL agent on one circuit and then apply the same agent to size new circuits or the same circuit in new technology nodes. In this way, we can reduce the simulation cost without designing from scratch.

Moreover, prior works such as Bayesian Optimization (BO) and Evolutionary Strategy (ES) treated transistor sizing as a black box optimization problem. Inspired by the simple fact that: \emph{circuit is a graph}, we propose to open the black box and leverage the \topology graph in the optimization loop. In order to make full use of the graph information, we propose to equip the RL agent with \emph{Graph Convolutional Neural Network} (GCN) to process the connection relationship between components in circuits. With the proposed GCN-RL agent, we consistently achieved better performance than conventional methods such as BO and ES. Remarkably, the GCN-RL not only enables transfer knowledge between different technology nodes but also makes knowledge transfer between \emph{different \topologies} possible. Experiments demonstrate that GCN is necessary for knowledge transfer between \topologies. 

To our knowledge, we are the first to leverage GCN equipped RL to transfer the knowledge between different technology nodes and different \topologies.
The contributions of this work are as follows:

\begin{enumerate}[(1)]
    \item \textbf{Leverage the Topology Graph Information} in the optimization loop (open-box optimization). We build a GCN based on the circuits \topology graph to effectively open the optimization black box and embed the domain knowledge of circuits to improve the performance.
    \item \textbf{Reinforcement Learning as Optimization Algorithm}, which consistently achieves better performance than human expert~\cite{stanford214Bhuman, stanford214Ahuman}, random search, Evolution Strategy (ES)~\cite{hansen2016cma}, Bayesian Optimization (BO)~\cite{snoek2012practical} and MACE~\cite{lyu2018batch}.
    \item \textbf{Knowledge Transfer with GCN-RL} between different technology nodes and different circuit \topologies to reduce the required number of simulations, thus shortening the design cycle.
\end{enumerate}

\begin{figure*}[t]
    \centering
    \includegraphics[width=0.9\linewidth]{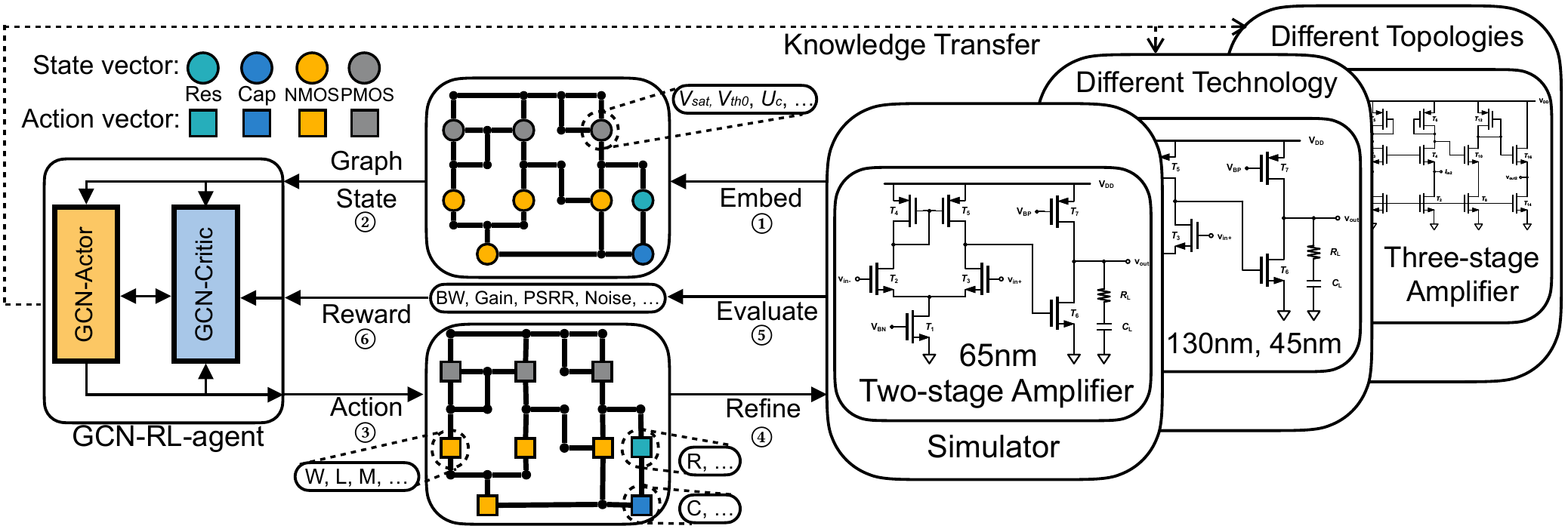}
    \vspace{-4pt}
    \caption{GCN-RL Overview: Six steps in one optimization iteration: (1) Embed \topology. (2) Generate graph and node states. (3) Generate actions (circuit parameters). (4) Refine circuit parameters. (5) Evaluate circuit parameters. (6) Update RL policy with the reward.
    }
    \vspace{-10pt}
    \label{fig:overview}
\end{figure*}

\section{Related work}
\myparagraph{Automatic Transistor Sizing}
Automatic transistor sizing can be classified into knowledge-based and optimization-based methods. For knowledge-based such as TAGUS~\cite{Horta2002},
circuits experts use their knowledge to design pre-defined plans and equations with which the transistor sizes are calculated. 
However, deriving a general design plan is highly time consuming, and it requires continuous maintenance to keep up with the latest technology. For optimization-based methods, they can be further categorized into model-based and simulation-based. Model-based methods such as~\cite{deniz2010hierarchical, castro2009multimode} model circuit performance via manual calculation or regression with simulated data, and then optimize the model. The advantage is the easy-to-get global optimal. Nevertheless, building a model requires numerous simulations to improve the accuracy. For simulation-based ones, performance of circuits is evaluated with simulators (\emph{e.g.} SPICE). The optimization algorithms such as BO~\cite{snoek2012practical}, MACE~\cite{lyu2018batch}, ES~\cite{hansen2016cma} consider the circuits as a black box and conduct optimization. Compared with ours, neither MACE nor ES leverage the \topology graph information. In addition, BO and MACE have difficulties in transferring knowledge between circuits because their output space is fixed. ES cannot transfer because it keeps good samples in its population without summarizing the design knowledge.

\myparagraph{Deep Reinforcement Learning}
Recently, deep RL algorithms have been extensively applied to many problems
such as game playing~\cite{silver2017mastering}, robotics~\cite{levine2016end} and AutoML~\cite{He:2018vj}. There are also environment libraries~\cite{mao2019park} using RL for system design.
For different RL task domains, deep RL proves to be transferable \cite{taylor2009transfer}. In this work, we propose RL based transistor sizing, which makes it  automated, transferable, and achieves better performance than other methods. Comparing to supervised learning, RL can continuously learn in the environment and adjust the policy.

\myparagraph{Graph Neural Networks}
Graph neural network (GNN) ~\cite{scarselli2009graph} adapts neural networks to process graph data. Several variants of GNN are proposed, including Graph Convolution Neural Networks (GCN)~\cite{kipf2016semi}, Graph Attention Networks \cite{velivckovic2017graph}, etc. There are also accelerators~\cite{yan2020hygcn, sparch} focusing on GNN related workloads.
In \NAME, we refer to~\cite{kipf2016semi} to build GCN, leveraging \topology graph information to benefit the optimization. \cite{circuitgnn} used GNN to replace an EM simulator for distributed circuits. By contrast, our work focuses on analog transistor sizing and exploits RL for knowledge transfer. 

\vspace{-5pt}
\section{Methodology}
\vspace{-5pt}
\subsection{Problem Definition}
We handle transistor sizing problem where the \topology of the analog circuits is \emph{fixed}. 
The problem can be formulated as a bound-constrained optimization:
\begin{equation}
\underset{x \in \mathbb{D}^{n}}{\text{maximize}} \quad \text{FoM}(x)
\label{fom:max}
\end{equation}
where $x$ is the parameter vector, $n$ is the number of parameters to search. $\mathbb{D}^{n}$ is the design space. Figure of Merits (FoM) is the objective we aim to optimize. We define it as the weighted sum of the normalized performance metrics as shown in Equation~\ref{fom:fom}.
\begin{equation}
\text{FoM} = \sum_{i=0}^N w_{i} \times \frac{\text{min}(m_{i}, m_{i}^{\text{bound}}) - m_{i}^\text{min}}{m_{i}^\text{max} - m_{i}^\text{min}} \text{\ if spec is satisfied}
\label{fom:fom}
\end{equation}
where $m_{i}$ is the measured performance metrics. $m_{i}^{\text{min}}$ and $m_{i}^{\text{max}}$ are pre-defined normalizing factors to normalize the performance metrics to guarantee their proper ranges. $m_{i}^{\text{bound}}$ is the pre-defined upper bound for some performance aspects which do not need to be better after satisfying some requirements. $w_{i}$ is the weight to adjust the importance of the $i^{\text{th}}$ performance metric. For some circuit baselines we use, there exists performance specification (spec) to meet, if the spec is not met, we assign a negative number as the FoM value.

\subsection{Framework Overview}
An overview of the proposed framework is shown in Figure~\ref{fig:overview}. 
In each iteration, (1) Circuit environment embeds the \topology into a graph whose vertices are components and edges are wires; (2) The environment generates a state vector for each transistor and passes the graph with the state vectors (refer to the graph on the top with circle nodes) to the RL agent; (3) The RL agent processes each vertex in the graph and generates an action vector for each node. Then the agent passes the graph with the node action vectors (refer to the graph with square vertices) to the environment; (4) The environment then denormalizes actions ([-1, 1] range) to parameters and refines them. We refine the transistor parameters to guarantee the transistor \emph{matching}. We also round and truncate parameters according to minimum precision, lower and upper bounds of the technology node; (5) Simulate the circuit; (6) Compute an FoM value and feed to RL agent to update policy. We \emph{do not need} the initial parameters as in the human design flow. The detailed RL agent will be discussed in Section~\ref{sec:gcnagent}.

\subsection{Reinforcement Learning Formulation}
We apply the actor-critic RL agent in GCN-RL. The critic can be considered as a differentiable model for the circuit simulator. The actor looks for points with best performance according to the model. 

\myparagraph{\textbf{State Space}}
The RL agent processes the circuit graph component by component. For a circuit with $n$ components in \topology graph $G$, the state $\mathbf{{s}_k}$ for the $k^{\text{th}}$ component is defined as $\mathbf{s_k} = (\mathbf{k}, \mathbf{t}, \mathbf{h})$,
where $\mathbf{k}$ is the one-hot representation of the transistor index, $\mathbf{t}$ is the one-hot representation of component type and $\mathbf{h}$ is the selected model feature vector for the component which further distinguishes different component types. For the NMOS and PMOS, the model parameters we use are $V_\text{sat}, V_\text{th0}, V_\text{fb}, \mu_{0}$ and $U_c$. For the capacitor and resistor, we set the model parameters to zeros.
For instance, for a circuit with ten components of four different kinds (NMOS, PMOS, R, C) and a five-dimensional model feature vector, the state vector for the third component (an NMOS transistor) is,
\begin{equation}
    [\color{orange}{0, 0, 1, 0, 0, 0, 0, 0, 0, 0}, \color{blue}{1, 0, 0, 0}, \color{violet}{V_\text{sat}, V_\text{th0}, V_\text{fb}, \mu_{0}, U_c}\color{black}{]}
\end{equation}
For each dimension in the observation vector $\mathbf{{s}_k}$, we normalize them by the mean and standard deviation across different components.

\myparagraph{\textbf{Action Space}}
\label{sect:approach:action_space}
The action vector varies for different types of components because the parameters needed to search are not the same. For the $k^{\text{th}}$ component, if it is NMOS or PMOS transistors, the action vector is formulated as $\mathbf{a_{k}^{MOS}} = (W, L, M)$, where $W$ and $L$ are the width and length of the transistor gate, $M$ is the multiplexer; for resistors, the action vector is formulated as: $\mathbf{a_{k}^{R}} = (r)$, where $r$ is the resistance value; for capacitors, the action vector is formulated as: $\mathbf{a_{k}^{C}} = (c)$, where $c$ is the capacitance value.

We use a \emph{continuous} action space to determine the transistor sizes even though we will round them to discrete values. The reason why we do not use a \emph{discrete} action space is because that will lose the \emph{relative order} information, also because the discrete action space is too large. For instance, for a typical operational amplifier with 20 transistors, each with three parameters, and each size with 1000 value options, the size of the discrete space is about $1000^{60}$.

\myparagraph{\textbf{Reward}}
The reward is the FoM defined in Equation~\ref{fom:fom}. It is a weighted sum of the normalized performance metrics. In our default setup, all the metrics are equally weighted. We also studied the effect of assigning different weights to different metrics in the experiments. Our method is \emph{flexible} to accommodate different reward setups. 
\label{sec:method}

\begin{figure}[t]
    \centering
    \includegraphics[width=\linewidth]{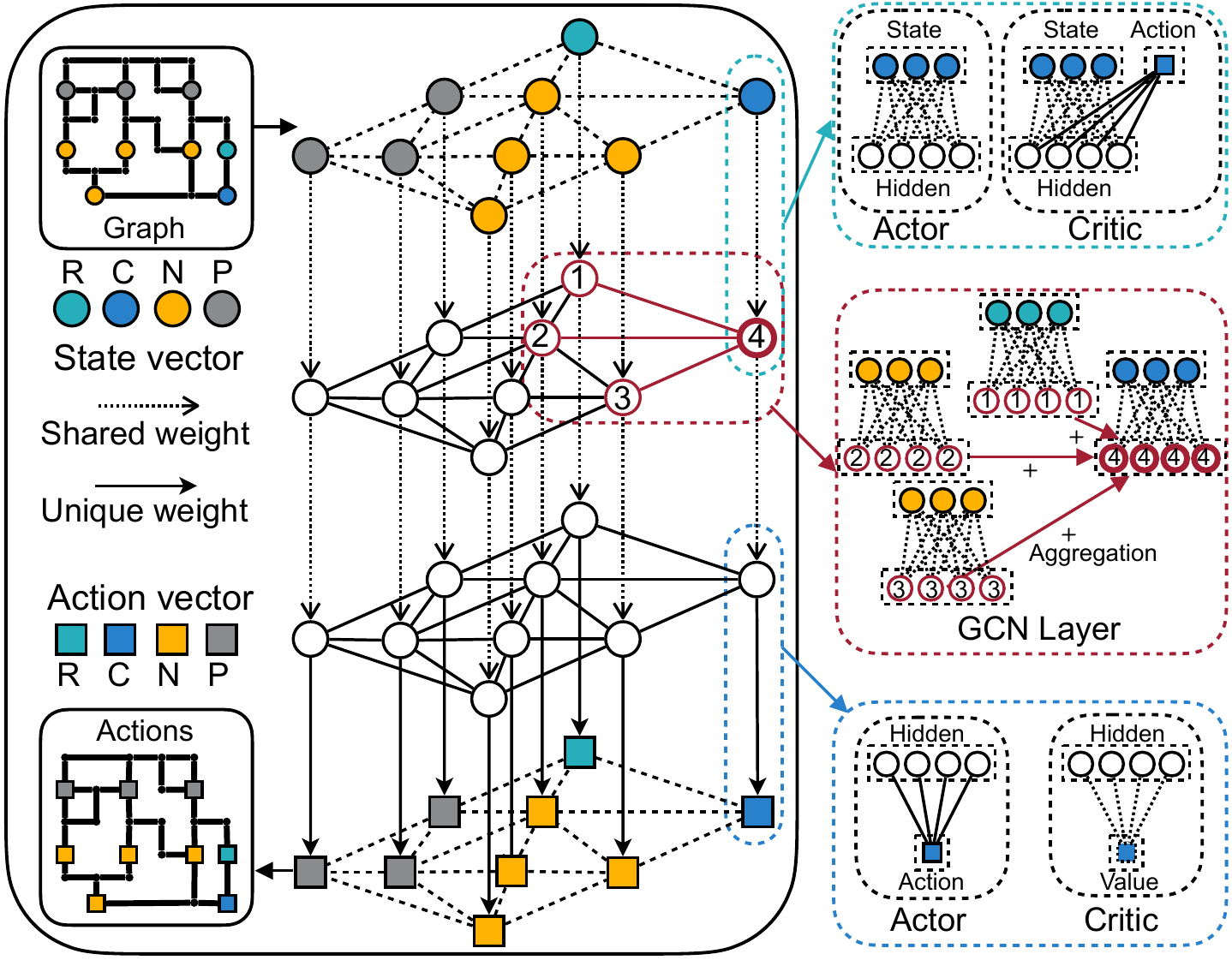}
    \caption{Reinforcement learning agent with multi-layer GCN.}

    \label{fig:gcn_arch}
    \vspace{-15pt}
\end{figure}

\subsection{Enhancing RL Agent with Graph Convolutional Neural Network}
\label{sec:gcnagent}

To embed the graph adjacency information into the optimization loop, we leverage GCN \cite{kipf2016semi} to process the \topology graph in the RL agent. As shown in Figure~\ref{fig:gcn_arch}, one GCN layer calculates each transistor's hidden representation by aggregating feature vectors from its neighbors. By stacking multiple layers, one node can receive information from farther and farther nodes. In our framework, we apply seven GCN layers to make sure the last layer have a \emph{global} receptive field over the entire \topology graph. The parameters for one component are primarily influenced by nearby components, but are also influenced by farther components. The GCN layer can be formulated as:
\begin{equation}
    H^{(l+1)} = \sigma(\widetilde{D}^{-\frac{1}{2}} \widetilde{A} \widetilde{D}^{-\frac{1}{2}}) H^{(l)}W^{(l)},
    \label{fom:gcn}
\end{equation}
Here, $\widetilde{A}=A+I_N$ is the adjacency matrix ($A$) of the \topology graph $G$  plus identity matrix ($I_N$). Adding the identity matrix is common in GCN networks\cite{kipf2016semi}. $\widetilde{D}_{ii} = \sum_{j} \widetilde{A}_{ij}$ and $W^{l}$ is a layer-specific trainable weight matrix, echoing with the shared weights in Figure~\ref{fig:gcn_arch}. $\sigma(\cdot)$ is an activation function such as ReLU($\cdot$)\cite{kipf2016semi}. $H^{(l)}\in \mathbb{R}^{n\times d}$ is the hidden features in the $l^{th}$ layer ($n$: number of nodes, $d$: feature dimension). $H^{0} = S$, which are the input state vectors for actor.

The actor and critic models have slightly different architectures (Figure~\ref{fig:gcn_arch}). The actor's first layer is a FC layer shared among all components. The critic's first layer is a shared FC layer with a \emph{component-specific} encoder to encode different actions. The actor's last layer has a \emph{component-specific} decoder to decode hidden activations to different actions, while the critic has a shared FC layer to compute the predicted reward value. We design those specific encoder/decoder layers because different components have \emph{different} kinds of actions (parameters). The output of the last layer of the actor is a pre-refined parameter vector for each component ranging [-1, 1]. We denormalize and refine them to get the final parameters. 
\begin{algorithm}[!t]
\SetKwInOut{Input}{Input}
    \SetAlgoLined
    Random Initialize critic network $Q(S, A \mid \theta^Q)$ and actor network $\mu(S \mid \theta^\mu)$ with critic weights $\theta^Q$ and actor weights $\theta^\mu$ \;
    Initialize replay buffer $P$ \; 
    \For{episode = 1, M}{
        Initialize random process $\mathcal{N}$;
        Receive observation state $S$\; 
        \If{episode $< W$ }{
            Warm-up: randomly sample an action $\widetilde{A}$\; 
        }
        \Else{
            Select action $\widetilde{A} = \mu(S \mid \theta^\mu) + \mathcal{N}$ according to the current policy and exploration noise\;
        }
        Denormalize and refine $\widetilde{A}$ with design constrains to get $A$\;
        Simulate the $A$ to get reward $R$\; 
        Store transition $(S, A, R)$ in $P$\;
        \If{episode $> W$}{
            Sample a batch of $(\widehat{S}, \widehat{A}, \widehat{R})$ from $P$ (batch size = $N_s$)\; 
            Update the critic by minimizing the loss:
            
            $L = \frac{1}{N_\text{s}}\sum_{k=1}^{N_\text{s}} (\widehat{R}_k - B - Q(\widehat{S}_k, \widehat{A}_k \mid \theta^Q))^2$\;
            
            Update the actor using the sampled policy gradient:
            
            \begin{footnotesize}
            $\nabla_{\theta^\mu}J \approx \frac{1}{N_\text{s}} \sum_{k=1}^{N_\text{s}} \nabla_{a}Q(S, A | \theta^Q) |_{\widehat{S}_{k}, \mu(\widehat{S}_{k})} \nabla_{\theta^\mu} \mu(S | \theta^\mu)|_{\widehat{S}_{k}}$\;
            \end{footnotesize}
        }
    }
    \caption{Proposed GCN-RL Method.}
    \label{algo}
\vspace{-3pt}
\end{algorithm}

For RL agent training, we leverage DDPG~\cite{Lillicrap:2016ww}, which is an off-policy actor-critic algorithm for continuous control problem. The details are illustrated in Algorithm~\ref{algo}. $N_\text{s}$ denotes the number of sampled data batch in one episode, $S=\{\mathbf{s_1, s_2, \ldots, s_n}\}$ denotes states, $A=\{\mathbf{a_1, a_2, \ldots, a_n}\}$ denotes actions. The baseline $B$ is defined as an exponential moving average of all previous rewards in order to \emph{reduce the variance} of the gradient estimation. $M$ is the max search episodes and $W$ is the warm-up episodes. $\mathcal{N}$ is a truncated norm noise with exponential decay.

We implemented two types of RL agent to show the effectiveness of GCN: one is the proposed GCN-RL and the other is non-GCN RL (NG-RL), which does not operate the \emph{aggregation} step in Figure~\ref{fig:gcn_arch}, thus \topology information is not used.

\vspace{-5pt}
\subsection{Knowledge Transfer}
\begin{figure}[t]
    \centering
    \vspace{-15pt}
    \includegraphics[width=\linewidth]{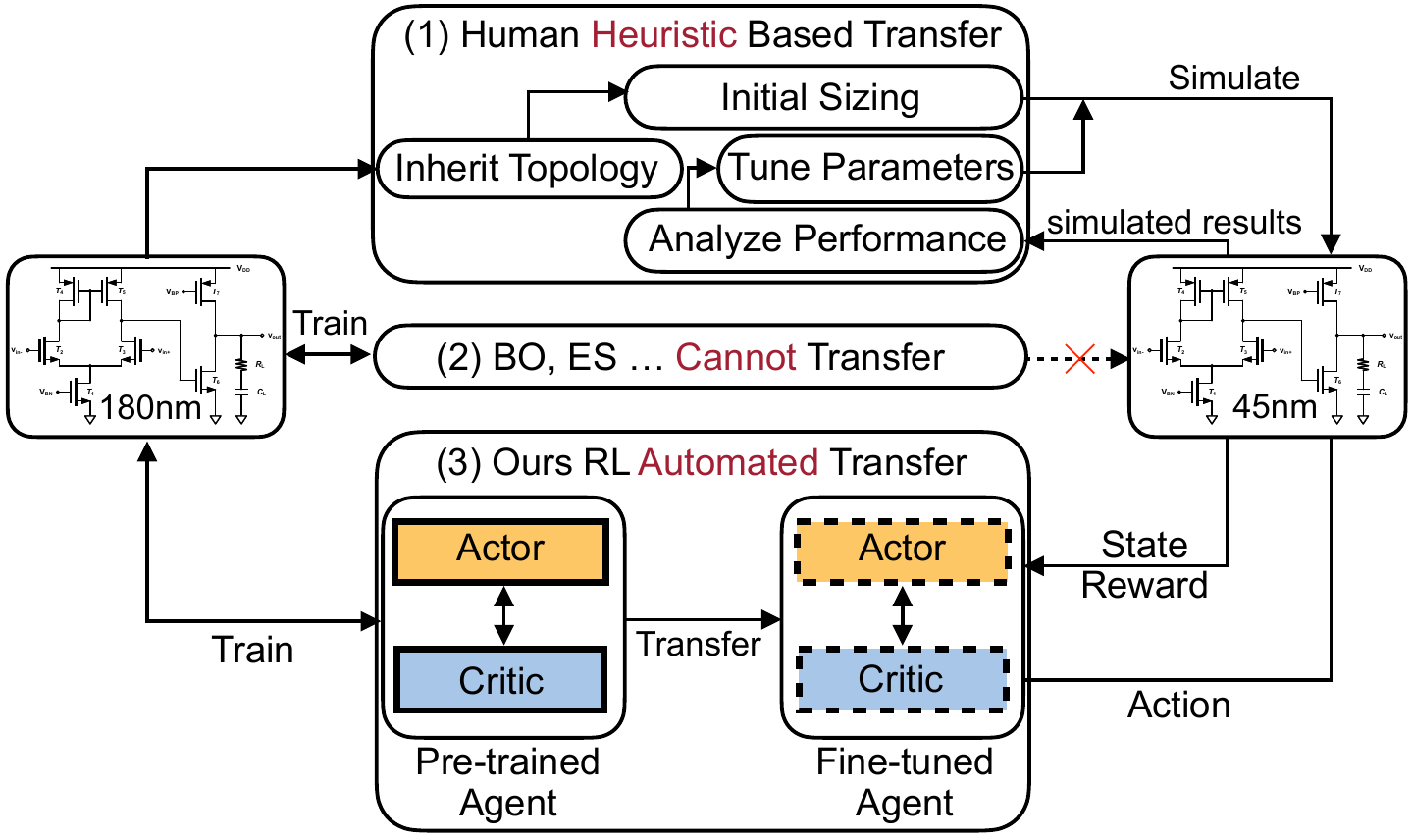}
    \vspace{-10pt}
    \caption{Our RL agent can transfer design knowledge by inheriting the pretrained weights of the actor-critic model. Human need use heuristic based methods to re-design the parameters. BO and ES cannot transfer.}
    \vspace{-10pt}
    \label{fig:transfer}
\end{figure}

\myparagraph{\textbf{Transfer between Technology Nodes}} The rapid transistor scaling down makes \emph{porting} existing designs from one technology node to another a common practice. As in Figure~\ref{fig:transfer} top, human designers first inherit the \topology from one node and compute initial parameters, then iteratively tune the parameters, simulate and analyze performance. In contrast, our method can automate this process by training an RL agent on one technology node and then \emph{directly applying} the trained agent to search the same circuit under different technology nodes by virtue of similar design principles among different technology. For instance, the agent can learn to change the gain by tuning the input pair transistors of an amplifier. This feature does not vary among technology nodes. 

\myparagraph{\textbf{Transfer between Topologies}} We can also transfer knowledge between different \topologies if they share similar design principles, such as between a two-stage transimpedance amplifier and a three-stage transimpedance amplifier. This is enabled by GCN which can extract features from the circuit \topologies. Concretely, we slightly modify the state vector mentioned in Section~\ref{sec:method}. $\mathbf{k}$ is modified to a one-dimension index value instead of a one-hot index vector. In this way, the dimension of the state vector of each component remains the \emph{same} among different \topologies.  We will show that without GCN, knowledge transfer between different \topologies cannot be achieved.  

% \vspace{-5pt}
\section{Experiments}
\subsection{Comparison between GCN-RL and others}
\label{sec:fomcomp}
\begin{figure*}
    \centering
    \vspace{-15pt}
    \includegraphics[width=1\textwidth]{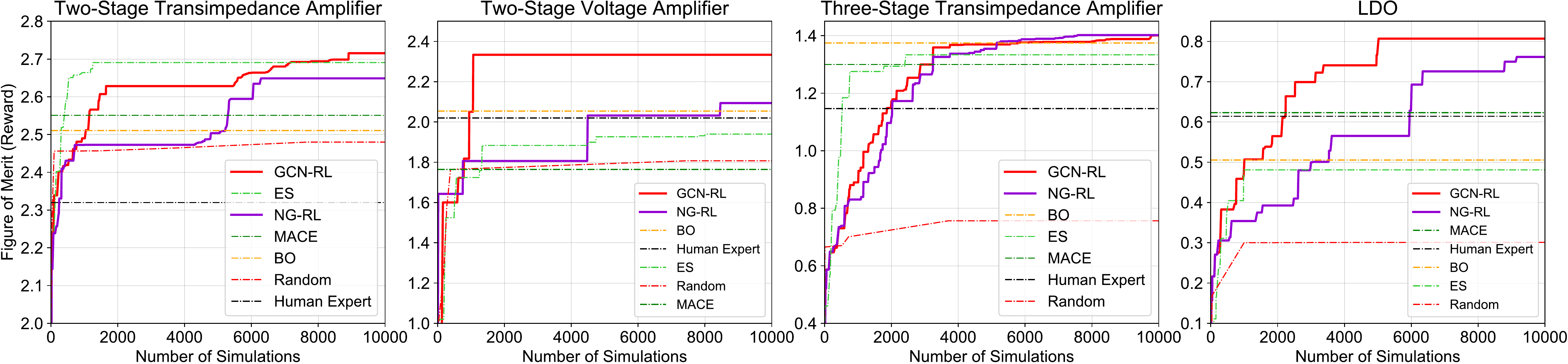}
    \vspace{-10pt}
    \caption{Learning curves of different algorithms. The proposed GCN-RL achieves the highest FoM in four baselines.}
    \vspace{-10pt}
    \label{fig:fom_comp}
\end{figure*}

\begin{figure*}
    \centering
    \includegraphics[width=1\textwidth]{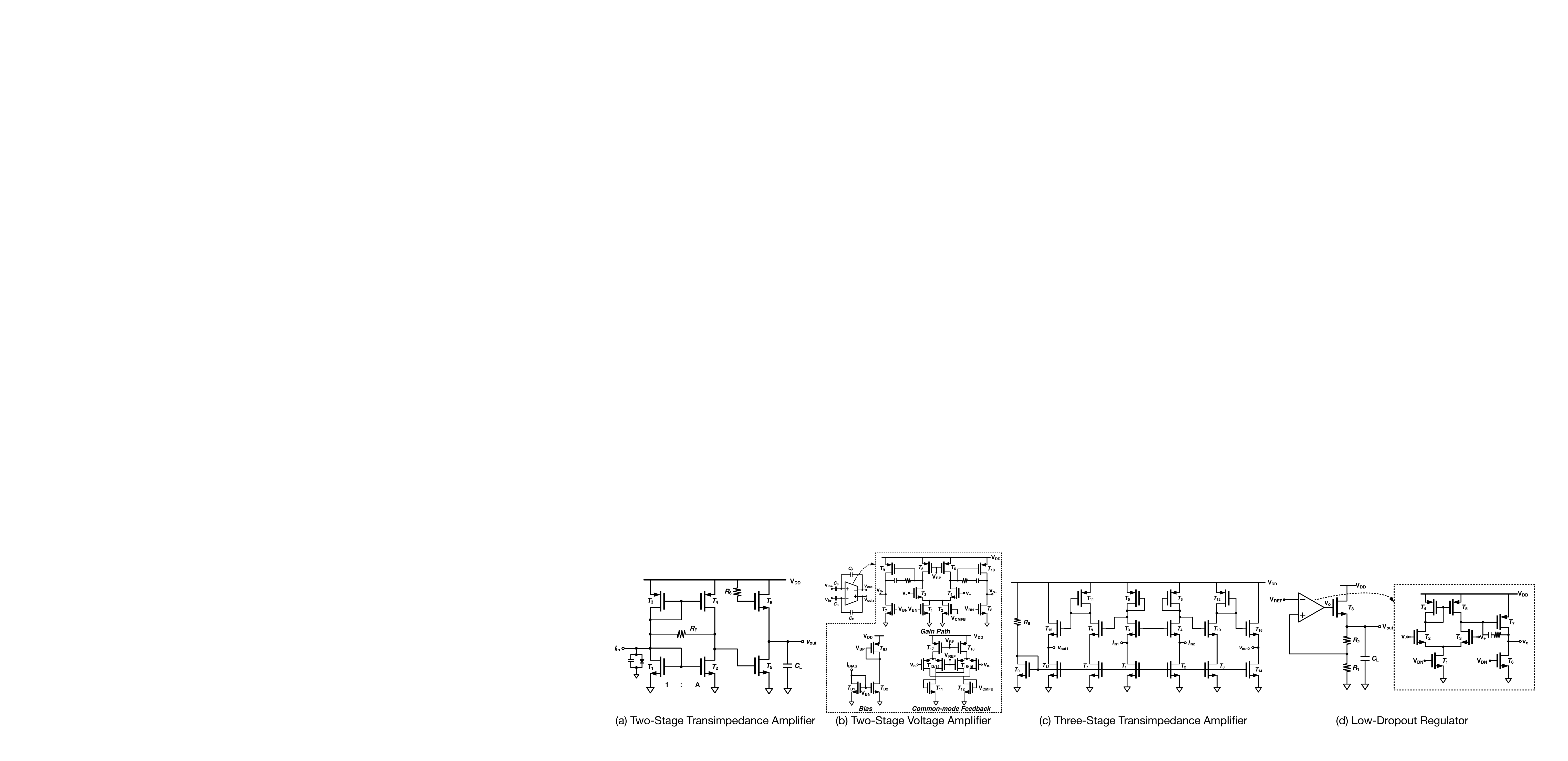}
    \vspace{-10pt}
    \caption{Topologies for four circuit baselines we optimize.}
    \vspace{-5pt}
    \label{fig:sche}
\end{figure*}

To demonstrate the effectiveness of the proposed GCN-RL method, we applied GCN-RL to 4 real-world circuits (Figure~\ref{fig:sche}): a two-stage transimpedance amplifier (Two-TIA), a two-stage voltage amplifier (Two-Volt), a three-stage transimpedance amplifier (Three-TIA) and a low-dropout regulator (LDO). The Two-Volt and LDO were designed in commercial 180nm TSMC technology, and simulated with Cadence Spectre. Two-TIA and Three-TIA were also designed in 180nm technology, simulated with Synopsys Hspice.

We compared FoMs of GCN-RL with human expert design, random search, non-GCN RL (NG-RL), Evolutionary Strategy (ES)~\cite{hansen2016cma}, Bayesian Optimization (BO)~\cite{snoek2012practical}, and MACE~\cite{lyu2018batch}. The human expert designs for Two-TIA and Three-TIA are strong baselines from~\cite{stanford214Bhuman, stanford214Ahuman}, which achieved the "Most Innovative Design" award~\cite{stanford214Bspec}. The human expert designs for Two-Volt and LDO come from a seasoned designer with 5-year experience designing for 6 hours. MACE is a parallel BO method with multi-objective acquisition ensemble. For ES, BO and MACE, we used open-sourced frameworks to guarantee the \emph{unbiased} implementations. For GCN-RL, NG-RL, ES and random search, we ran 10000 steps. In these methods, the circuit simulation time accounts for over 95\% of the total runtime. For BO and MACE, it is \emph{impossible} to run 10000 steps because the computation complexity is $\mathcal{O}(N^3)$, thus we ran them for the \emph{same runtime} (around 5 hours) with GCN-RL for fair comparisons.

We ran each experiment three times to show significance. We computed the FoM values based on Equation~\ref{fom:fom}. The $m_i^\text{max}$ and $m_i^\text{min}$ were obtained by random sampling 5000 designs, and choosing the max and min of each metric. We assigned $w_i=1$ to one performance metric if larger is better, such as gain; and assigned $w_i=-1$ if smaller is better, such as power. The experimental results are shown in Table~\ref{tab:part1}. We plot the max FoM values among three runs for each method in Figure~\ref{fig:fom_comp}. GCN-RL consistently achieves the highest FoM. The \emph{convergence speed} of GCN-RL is also faster than NG-RL, which benefits from the fact that GCN is better than pure FC on extracting the \topology features, similar to CNN being better than pure FC on extracting image features.

\myparagraph{\textbf{Two-Stage Transimpedance Amplifier}}
Diode-connected input transistors are used to convert its drain current to a voltage at the drain of its output stage. Bandwidth, gain, power, noise and peaking are selected as metrics. The FoM results are shown in the `Two-TIA' column of Table~\ref{tab:part1}. Performance metrics are show in Table~\ref{tab:twotrans} top part. All the models satisfy the spec. Since the FoM encourages \emph{balanced} performance among metrics, the GCN-RL design forms a good balance among five metrics, resulting in the highest FoM. Notably, GBW of GCN-RL is also the highest. 

We often need to trade-off different performance metrics. In order to demonstrate the \emph{flexibility} of the GCN-RL with different design focuses, we assigned 10$\times$ larger weight for one metric then other metrics: for GCN-RL-1 to GCN-RL-5, 10$\times$ larger weight on each of BW, gain, power, noise and peaking. Since one spec is only suitable for one design requirement, in those five experiments, we \emph{did not} limit the results with the spec to show its general effectiveness. Except for GCN-RL-4, all designs achieve highest performance on the single aspect we care about. For GCN-RL-4, it achieves the second best.

\begin{table}[t]
\setlength{\tabcolsep}{2pt}
    \centering
   \begin{footnotesize}
    \begin{tabular}{lcccc}
    \toprule
         & Two-TIA & Two-Volt  & Three-TIA & LDO \\
         \hline
         Human~\cite{stanford214Ahuman, stanford214Bhuman}   & 2.32  & 2.02                       & 1.15            & 0.61 \\
         \hline
         Random   & 2.46 $\pm$ 0.02 & 1.74 $\pm$ 0.06  & 0.74 $\pm$ 0.03 & 0.27 $\pm$ 0.03 \\
         \hline
         ES~\cite{hansen2016cma}      & 2.66 $\pm$ 0.03 & 1.91 $\pm$ 0.02  & 1.30 $\pm$ 0.03 & 0.40 $\pm$ 0.07 \\
         \hline
         BO~\cite{snoek2012practical}  & 2.48 $\pm$ 0.03     & 1.85 $\pm$ 0.19  & 1.24 $\pm$ 0.14 & 0.45 $\pm$ 0.05 \\
         \hline
         MACE~\cite{lyu2018batch}   & 2.54 $\pm$ 0.01  & 1.70 $\pm$ 0.08  & 1.27 $\pm$ 0.04 & 0.58 $\pm$ 0.04 \\
         \hline
         \textbf{NG-RL} & 2.59 $\pm$ 0.06 & 1.98 $\pm$ 0.12  & 1.39 $\pm$ 0.01 & 0.71 $\pm$ 0.05 \\
         \hline
  \textbf{GCN-RL} &\textbf{ 2.69 $\pm$ 0.03 } & \textbf{  2.23 $\pm$ 0.11}  & \textbf{ 1.40 $\pm$ 0.01} & \textbf{0.79 $\pm$ 0.02} \\
    \bottomrule
    \end{tabular}
    \end{footnotesize}
  
    \caption{FoM comparison between different algorithms. Our method consistently achieves the highest FoM values in four circuits baselines.}
    \label{tab:part1}
    \vspace{-10pt}
\end{table}

\begin{table}[t]

\setlength{\tabcolsep}{2pt}
    \centering
    \begin{footnotesize}
    \begin{tabular}{lcccccccc}
    \toprule
         & BW  & Gain & Power & Noise  & Peaking & GBW &FoM \\
         & (GHz) & ($\times10^{2}\Omega$) & (mW) & (pA/$\sqrt{\text{Hz}}$) & (dB) & (THz$\times \Omega$)\\
         \hline
         
  Spec~\cite{stanford214Bspec} 
  & max & $>$7.58 & $<$18.0 & $<$19.3 & $<$1.00 & max & -\\
  \hline
  Human~\cite{stanford214Bhuman}  & 5.95 & 7.68 & 8.11 & 18.6 & 0.93 & 4.57 & 2.32\\
  \hline
  Random & 1.41 & 35.9 & 2.30 & 9.28 & 0.16 & 5.05 & 2.48\\
  \hline
  ES~\cite{hansen2016cma} & 1.18 & 104 & 4.25 & 3.77 & 0 & 12.3 & 2.69\\
  \hline
  BO~\cite{snoek2012practical}     & 0.16 & 123 & 2.76 & \textbf{1.68} & 0 & 1.99 & 2.51\\
  \hline
  MACE~\cite{lyu2018batch}   & 0.97 & 83.1 & 2.74 & 7.36 & 0 & 8.07 & 2.55 \\
  \hline
\textbf{NG-RL} & 0.75 & 156 & 2.31 & 3.85 & 0.068 & 11.7 & 2.65 \\ \hline
\textbf{GCN-RL}   & 1.03 & 167 & 3.44 & 3.72 & 0.0003 & \textbf{17.2} & \textbf{2.72} \\ \hline \hline
\textbf{GCN-RL-1} & \textbf{13.6} & 0.09  & 41.2 & 295 & 0.036 & 0.13 & - \\ \hline
\textbf{GCN-RL-2} & 0.20  & \textbf{266}  & 2.58 & 5.73 & 0  & 5.18 & - \\ \hline
\textbf{GCN-RL-3} & 0.42  & 249  & \textbf{0.58} & 4.78 & 0  & 10.3 & - \\ \hline
\textbf{GCN-RL-4} & 0.86  & 124  & 3.67 & \textbf{3.64} & 1.0  & 10.7 & - \\ \hline
\textbf{GCN-RL-5} & 0.57  & 89.0  & 0.94 & 11.7 & \textbf{0}  & 5.10 & - \\

    \bottomrule
    \end{tabular}
    \end{footnotesize}
    
    \caption{Performance metrics comparison of two-stage transimpedance amplifier. First 8 rows: GCN-RL forms a good balance between different metrics and achieves highest GBW and FoM. Last 5 rows: GCN-RL is \emph{flexible} for different FoMs. It can achieve best single-metric performance if the FoM value is biased to one performance metric.}
    \vspace{-10pt}
    \label{tab:twotrans}
\end{table}

\myparagraph{\textbf{Two-Stage Voltage Amplifier}}
The amplifier is connected in a closed-loop configuration for PVT-stable voltage gain, which is set by the capacitor ratio. Miller compensation is used to stabilize the closed-loop amplifier. Bandwidth, common mode phase margin (CPM), differential mode phase margin (DPM), power, noise and open loop gain are selected as metrics. The FoM results are shown in the `Two-Volt' column of Table~\ref{tab:part1} and details in Table~\ref{tab:twodiff}. The GCN-RL achieves highest CPM, DPM; second highest Gain and GBW.

\begin{table}[t]
\setlength{\tabcolsep}{2pt}
    \centering
    \begin{footnotesize}
    \begin{tabular}{lcccccccc}
    \toprule
         & BW  & CPM & DPM & Power & Noise & Gain & GBW & FoM \\
        
         & (MHz) & ($\degree$) & ($\degree$) & ($\times10^{-4}$W) & (nA/$\sqrt{\text{Hz}}$) & ($\times 10^3$) & (THz) &  \\
         \hline
  Human  & \textbf{242} & 180 & 83.9 & 2.94 & 47.1 & 3.94 & 0.95 & 2.02 \\
  \hline
  Random & 187 & 180 & 2.51 & 7.85 & 23.8 & 8.77 & 1.64 & 1.80 \\
  \hline
  ES~\cite{hansen2016cma}     & 27.3 & 180 & 4.43 & 1.46 & 74.2 & \textbf{50.0} & 1.37 & 1.93\\
  \hline
  BO~\cite{snoek2012practical}     & 151 & 166 & 2.77 & \textbf{1.45} & 46.3 & 25.0 & \textbf{3.77} & 2.05\\
  \hline
  MACE~\cite{lyu2018batch} & 99.4 & 180 & 4.44 & 8.45 & \textbf{16.1} & 8.93 & 0.89 & 1.76 \\
  \hline
\textbf{NG-RL} & 96.2 & 180 & 23.7 & 4.02 & 19.2 & 17.2 & 1.66  & 2.09\\
\hline
\textbf{GCN-RL} & 84.7 & \textbf{180} & \textbf{96.3} & 2.56 & 58.7 & 29.4 & 2.57 & \textbf{2.33} \\

    \bottomrule
    \end{tabular}
    \end{footnotesize}
    \caption{Performance metrics comparison of two-stage voltage amplifier. GCN-RL forms a good balance between metrics and achieves best CPM and DPM and second highest Gain and GBW.}
    \vspace{-10pt}
    \label{tab:twodiff}
\end{table}

\myparagraph{\textbf{Three-Stage Transimpedance Amplifier}}
A common-mode input pairs is used to convert its differential source current to a voltage at the drain of its output stage. Three-stage configuration is used to boost the I-V conversion gain. Bandwidth, gain and power are selected as metrics. The FoM results are shown in the `Three-TIA' column of Table~\ref{tab:part1}. GCN-RL achieves lowest power and second highest GBW.

\begin{figure*}[t]
    \centering
    \vspace{-5pt}
    \includegraphics[width=\textwidth]{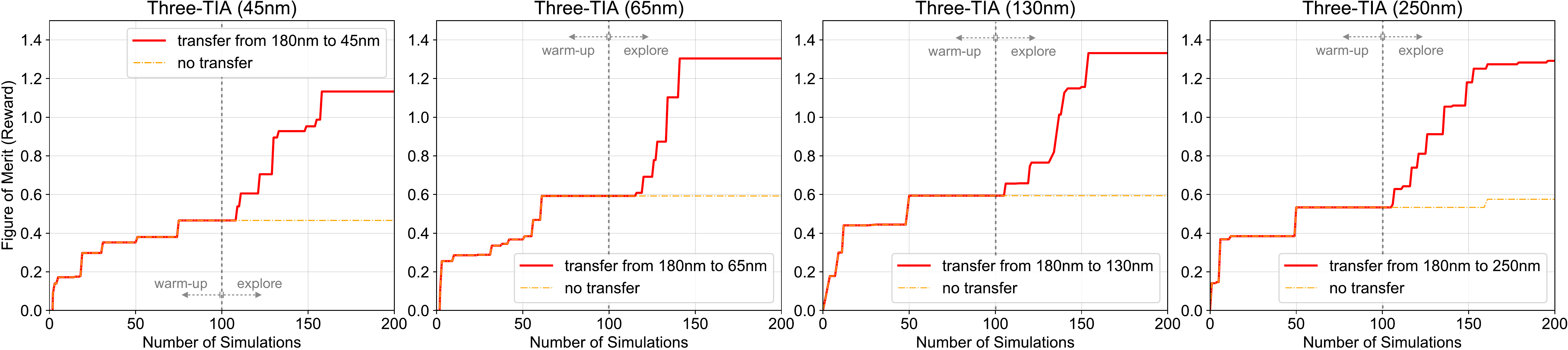}
    \vspace{-10pt}
    \caption{Knowledge transfer between technology nodes on Three-TIA. We transfer the design learned on 180nm to 45nm, 65nm, 130nm, and 250nm. After 100 warm-up steps, the FoM with knowledge transfer increases rapidly and finally converges at a higher level than without knowledge transfer. }
    \vspace{-8pt}
    \label{fig:threetranstransfer}
\end{figure*}

\myparagraph{\textbf{Low-Dropout Regulator}}
LDO regulates the output voltage and minimizes the effect of either supply change or load change. 
Settling time for load increase ($T_{L+}$), settling time for load decrease ($T_{L-}$), Load Regulation (LR), settling time for supply increase ($T_{V+}$), settling time for supply decrease ($T_{V+}$), Power Supply Rejection Ratio (PSRR) and power are the metrics. The FoM results are shown in the `LDO' column of Table~\ref{tab:part1}. The proposed GCN-RL method achieves the shortest $T_{L+}$ and $T_{L-}$, second largest LR and PSRR.

\vspace{-5pt}
\subsection{Knowledge Transfer Between Technology Nodes}
We also conducted experiments on knowledge transfer between technology nodes, which is a great feature enabled by reinforcement learning. The agents are trained on 180nm and transferred to both larger node -- 250nm and smaller nodes -- 130, 65 and 45nm to verify its extensive effectiveness. 

\myparagraph{Transfer on Two-Stage Transimpedance Amplifier}
We directly applied the RL agent trained in Section~\ref{sec:fomcomp} on searching parameters for Two-TIA on other technology nodes. We compared the transfer learning results with no transfer after limited number of training steps (300 in total: 100 warm-up, 200 exploration). As in Table~\ref{tab:twostagetechtrans} top part,
transfer learning results are much better than without transfer.

\begin{table}[t]
\setlength{\tabcolsep}{1.8pt}
\begin{footnotesize}
    \centering
    \begin{tabular}{ccccc}
    \toprule
                        & 250nm & 130nm & 65nm & 45nm \\

\hline
Two-TIA   & \multirow{2}{*}{2.36 $\pm$ 0.05} & \multirow{2}{*}{2.43 $\pm$ 0.03} & \multirow{2}{*}{2.36 $\pm$ 0.09} & \multirow{2}{*}{2.36 $\pm$ 0.06} \\

No Transfer\\
\hline
Two-TIA  &\textbf{\multirow{2}{*}{2.55 $\pm$ 0.01}} & \textbf{\multirow{2}{*}{2.56 $\pm$ 0.02}} & \textbf{\multirow{2}{*}{2.52 $\pm$ 0.04}} & \textbf{\multirow{2}{*}{2.51 $\pm$ 0.04}} \\
Transfer from 180nm\\
\hline
\hline
Three-TIA   &  \multirow{2}{*}{0.69 $\pm$ 0.25} &  \multirow{2}{*}{0.65 $\pm$ 0.14} &  \multirow{2}{*}{0.55 $\pm$ 0.03} &  \multirow{2}{*}{0.53 $\pm$ 0.05} \\
No Transfer\\
\hline
Three-TIA  &  \textbf{\multirow{2}{*}{1.27 $\pm$ 0.02}} &  \textbf{\multirow{2}{*}{1.29 $\pm$ 0.05}} &  \textbf{\multirow{2}{*}{1.20 $\pm$ 0.09}} &  \textbf{\multirow{2}{*}{1.06 $\pm$ 0.07}} \\
Transfer from 180nm\\
\bottomrule
    \end{tabular}
    \end{footnotesize}
    \caption{Knowledge transfer from 180nm to other technology nodes on Two-TIA and Three-TIA. After the \emph{same} steps, performance with transfer is \emph{consistently better} than without transfer.}
    \vspace{-10pt}
    \label{tab:twostagetechtrans}
\end{table}

\myparagraph{Transfer on Three-Stage Transimpedance Amplifier}
We also applied the RL agent in Section~\ref{sec:fomcomp} to search the parameters on other nodes. We show the results in Table~\ref{tab:twostagetechtrans} bottom part. We also plot the max FoM value learning curves in Figure~\ref{fig:threetranstransfer}. We use the same random seeds for two methods so they have the same FoMs in the warm-up stage. After exploration, transfer learning results are consistently better than no transfer after the \emph{same} steps.

\subsection{Knowledge Transfer Between Topologies}
\begin{figure}
    \centering
    \includegraphics[width=1\columnwidth]{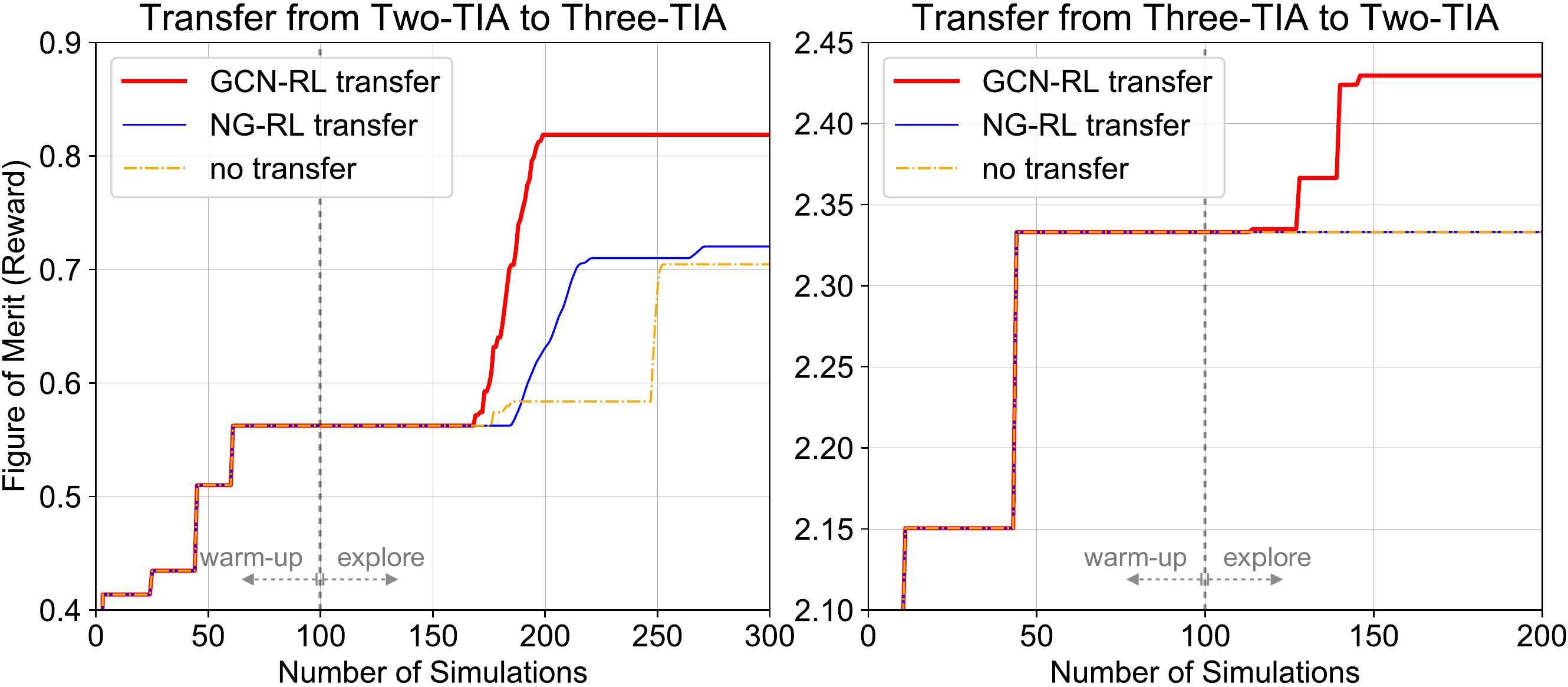}
    \vspace{-10pt}
    \caption{Knowledge transfer between Two-TIA and Three-TIA. After 100 warm-up steps, the FoM of GCN-RL increases rapidly and finally converges at a higher value than NG-RL transfer and no transfer.}
    \vspace{-8pt}
    \label{fig:transfersch}
\end{figure}

We can also transfer the knowledge learned from one \topology to another. We chose Two-TIA and Three-TIA as they are both transimpedance amplifier, thus sharing some common knowledge. 
We first trained both GCN-RL and NG-RL agents on Two-TIA for 10000 steps. Then, we directly applied the agents to the Three-TIA and trained for only 300 steps.
We also conducted the reverse experiments, learning from Three-TIA and transferring to Two-TIA. We compared (1) transfer learning with GCN-RL, (2) transfer learning without GCN (NG-RL), (3) no transfer with GCN-RL, as in Table~\ref{tab:transfersch}, and learning curves in Figure~\ref{fig:transfersch}. GCN-RL consistently achieves higher FoMs than NG-RL. Without GCN, the FoM of NG-RL is barely at the \emph{same} level as without transfer, which shows that it is \emph{critical} to use GCN to extract the knowledge from the graph, and the graph information extracted by GCN can help improve knowledge transfer performance.

\begin{table}[t]
    \centering
    \setlength{\tabcolsep}{2pt}
    \begin{footnotesize}
    \begin{tabular}{lccc}
    \toprule
         & Two-TIA $\xrightarrow{}$ Three-TIA & Three-TIA $\xrightarrow{}$ Two-TIA \\
         \hline
        No Transfer & 0.63 $\pm$ 0.07 & 2.37 $\pm$ 0.01 \\
        \hline
        NG-RL Transfer & 0.62 $\pm$ 0.09 & 2.40 $\pm$ 0.07  \\
        \hline
        GCN-RL Transfer & \textbf{0.78 $\pm$ 0.12} & \textbf{2.45 $\pm$ 0.02} \\
    \bottomrule
    \end{tabular}
    \end{footnotesize}
    \caption{Knowledge transfer between Two-TIA and Three-TIA. GCN-RL consistently achieves higher FoM than NG-RL. NG-RL transfer has similar FoM as no transfer.}
    \vspace{-10pt}
    \label{tab:transfersch}
\end{table}

\section{Conclusion}
We present \NAME, a transferable automatic transistor sizing method with GCN and RL. 
Benefiting from transferability of RL, we can transfer the knowledge between different technology nodes and even different \topologies, which is difficult with other methods.
We also use GCN to involve \topology information into the RL agent. Extensive experiments demonstrate that our method can achieve better FoM than others, with knowledge transfer ability. 
Therefore, \NAME enables more effective and efficient transistor sizing and design porting.

\section*{Acknowledgment}
\vspace{5pt}
\small{
We thank NSF Career Award \#1943349, MIT Center for Integrated Circuits and Systems, Samsung, MediaTek for supporting this research.
}

\vspace{5pt}

\bibliographystyle{IEEEtran}
\bibliography{main.bib}

\begin{thebibliography}{10}
\providecommand{\url}[1]{#1}
\csname url@samestyle\endcsname
\providecommand{\newblock}{\relax}
\providecommand{\bibinfo}[2]{#2}
\providecommand{\BIBentrySTDinterwordspacing}{\spaceskip=0pt\relax}
\providecommand{\BIBentryALTinterwordstretchfactor}{4}
\providecommand{\BIBentryALTinterwordspacing}{\spaceskip=\fontdimen2\font plus
\BIBentryALTinterwordstretchfactor\fontdimen3\font minus
  \fontdimen4\font\relax}
\providecommand{\BIBforeignlanguage}[2]{{%
\expandafter\ifx\csname l@#1\endcsname\relax
\typeout{** WARNING: IEEEtran.bst: No hyphenation pattern has been}%
\typeout{** loaded for the language `#1'. Using the pattern for}%
\typeout{** the default language instead.}%
\else
\language=\csname l@#1\endcsname
\fi
#2}}
\providecommand{\BIBdecl}{\relax}
\BIBdecl

\bibitem{elfadel2018machine}
I.~A.~M. Elfadel \emph{et~al.}, \emph{Machine Learning in VLSI Computer-Aided
  Design}.\hskip 1em plus 0.5em minus 0.4em\relax Springer, 2018.

\bibitem{lyu2018batch}
W.~Lyu \emph{et~al.}, ``Batch bayesian optimization via multi-objective
  acquisition ensemble for automated analog circuit design,'' in \emph{ICML},
  2018.

\bibitem{liao2017parasitic}
T.~Liao \emph{et~al.}, ``Parasitic-aware gp-based many-objective sizing
  methodology for analog and rf integrated circuits,'' in \emph{ASP-DAC}, 2017.

\bibitem{liu2010enhanced}
B.~Liu \emph{et~al.}, ``An enhanced moea/d-de and its application to
  multiobjective analog cell sizing,'' in \emph{CEC}, 2010.

\bibitem{learncircuits}
H.~Wang \emph{et~al.}, ``Learning to design circuits,'' in \emph{NeurIPS
  Machine Learning for Systems Workshop}, 2018.

\bibitem{stanford214Bhuman}
D.~Bankman. (2013) {Stanford EE214B Advanced Analog IC Design Contest "Most
  Innovative Design Award". (Available upon request).}

\bibitem{stanford214Ahuman}
D.~Bankman \emph{et~al.} (2013) {Stanford EE214A Fundamentals of Analog IC
  Design, Design Project Report. (Available upon request).}

\bibitem{hansen2016cma}
N.~Hansen, ``The cma evolution strategy: A tutorial,'' \emph{arXiv preprint
  arXiv:1604.00772}, 2016.

\bibitem{snoek2012practical}
J.~Snoek \emph{et~al.}, ``Practical bayesian optimization of machine learning
  algorithms,'' in \emph{NIPS}, 2012.

\bibitem{Horta2002}
N.~Horta, ``Analogue and mixed-signal systems topologies exploration using
  symbolic methods,'' \emph{AICSP}, 2002.

\bibitem{deniz2010hierarchical}
E.~Deniz \emph{et~al.}, ``Hierarchical performance estimation of analog blocks
  using pareto fronts,'' in \emph{6th Conference on Ph.D. Research in
  Microelectronics Electronics}, 2010.

\bibitem{castro2009multimode}
R.~Castro-L{\'o}pez \emph{et~al.}, ``Multimode pareto fronts for design of
  reconfigurable analogue circuits,'' \emph{Electronics Letters}, 2009.

\bibitem{silver2017mastering}
D.~Silver \emph{et~al.}, ``Mastering the game of go without human knowledge,''
  \emph{Nature}, 2017.

\bibitem{levine2016end}
S.~Levine \emph{et~al.}, ``End-to-end training of deep visuomotor policies,''
  \emph{JMLR}, 2016.

\bibitem{He:2018vj}
Y.~He \emph{et~al.}, ``{AMC: AutoML for Model Compression and Acceleration on
  Mobile Devices},'' in \emph{ECCV}, 2018.

\bibitem{mao2019park}
H.~Mao \emph{et~al.}, ``Park: An open platform for learning-augmented computer
  systems,'' in \emph{NeurIPS}, 2019.

\bibitem{taylor2009transfer}
M.~E. Taylor \emph{et~al.}, ``Transfer learning for reinforcement learning
  domains: A survey,'' \emph{JMLR}, 2009.

\bibitem{scarselli2009graph}
F.~Scarselli \emph{et~al.}, ``The graph neural network model,'' \emph{IEEE
  Transactions on Neural Networks}, vol.~20, no.~1, pp. 61--80, 2009.

\bibitem{kipf2016semi}
T.~N. Kipf \emph{et~al.}, ``Semi-supervised classification with graph
  convolutional networks,'' \emph{arXiv preprint arXiv:1609.02907}, 2016.

\bibitem{velivckovic2017graph}
P.~Veli{\v{c}}kovi{\'c} \emph{et~al.}, ``Graph attention networks,''
  \emph{arXiv preprint arXiv:1710.10903}, 2017.

\bibitem{yan2020hygcn}
M.~Yan \emph{et~al.}, ``Hygcn: A gcn accelerator with hybrid architecture,'' in
  \emph{HPCA}.\hskip 1em plus 0.5em minus 0.4em\relax IEEE, 2020.

\bibitem{sparch}
Z.~Zhang \emph{et~al.}, ``Sparch: Efficient architecture for sparse matrix
  multiplication,'' in \emph{HPCA}.\hskip 1em plus 0.5em minus 0.4em\relax
  IEEE, 2020.

\bibitem{circuitgnn}
G.~Zhang \emph{et~al.}, ``Circuit-{GNN}: Graph neural networks for distributed
  circuit design,'' in \emph{ICML}, 2019.

\bibitem{Lillicrap:2016ww}
T.~Lillicrap \emph{et~al.}, ``{Continuous control with deep reinforcement
  learning.}'' in \emph{ICLR}, 2016.

\bibitem{stanford214Bspec}
\BIBentryALTinterwordspacing
B.~Murmann. (2013) Stanford ee214b advanced analog ic design contest. [Online].
  Available: \url{https://web.stanford.edu/class/ee214b/contest}
\BIBentrySTDinterwordspacing

\end{thebibliography}

\end{document}